\documentclass[a4paper,11pt]{article}
\usepackage[dvipsnames]{xcolor}
\usepackage[utf8]{inputenc}
\usepackage[bbgreekl]{mathbbol}
\usepackage{geometry}
\usepackage{multicol}

\usepackage{extarrows}
\usepackage{xcolor}
\usepackage{amsmath, array, amssymb, amsfonts,amsthm}
\usepackage{upgreek}
\usepackage{xfrac}
\usepackage[inline]{enumitem}
\setlist{nolistsep}
\usepackage[french, italian, english]{babel}
\usepackage[all]{xy}
\usepackage{txfonts}  
\usepackage{sectsty}
\usepackage{booktabs}
\usepackage{caption}
\usepackage{dsfont}
\usepackage{mathtools}
\usepackage{slashed}
\usepackage[makeroom]{cancel}
\usepackage[hidelinks]{hyperref}
\usepackage{textcomp}
\usepackage{multicol}
\usepackage[title]{appendix}
\usepackage[square,numbers,sort,compress,merge]{natbib}
\let\cite\citep 
\usepackage{hyperref}
\hypersetup{colorlinks=true, urlcolor=blue, citecolor=blue, linktoc=page}

\usepackage{tikz}
\usepackage{tikz-cd}
\usetikzlibrary{cd}
\tikzcdset{
arrow style=tikz,
diagrams={>={Straight Barb[scale=0.8]}}
}
\usetikzlibrary{matrix,arrows,decorations.pathmorphing}

\DeclareMathAlphabet{\mathpzc}{OT1}{pzc}{m}{it}

\usepackage{fancybox}

\usepackage{mathrsfs}
\usepackage{scrextend}

\makeatletter
\renewcommand*\env@matrix[1][\arraystretch]{%
  \edef\arraystretch{#1}%
  \hskip -\arraycolsep
  \let\@ifnextchar\new@ifnextchar
  \array{*\c@MaxMatrixCols c}}
\makeatother

\newcommand{\defeq}{\vcentcolon=}
\newcommand{\rdefeq}{=\vcentcolon}

\newcommand\RR{\mathbb{R}}
\newcommand\CC{\mathbb{C}}

\renewcommand\1{\textbf{1}}
\newcommand\id{\textit{id}}
\newcommand\T{\mathcal{T}}
\newcommand\G{\mathcal{G}}
\renewcommand\H{\mathcal{H}}

\newcommand\SU{\mathcal{SU}}

\newcommand\SO{\mathcal{SO}}
\newcommand\SL{\mathcal{SL}}

\newcommand\K{\mathcal{K}}
\newcommand\J{\mathcal{J}}

\newcommand\GL{\mathcal{GL}}

\newcommand\vphi{\varphi}

\renewcommand\epsilon{\varepsilon}

\newcommand\rarrow{\rightarrow}

\newcommand\LieG{\mathfrak{g}}
\newcommand\LieH{\mathfrak{h}}

\renewcommand\t{\tilde}

\renewcommand\b{\bar }

\renewcommand\-{^{-1}}
\newcommand\Ad{\text{Ad}}

\renewcommand\id{\text{id}}
\renewcommand\1{\mathds{1}}

\makeatletter

\newcommand{\Rmnum}[1]{\expandafter\@slowromancap\romannumeral #1@}
\makeatother

\makeatletter
\newcommand{\leqnomode}{\tagsleft@true\let\veqno\@@leqno}
\newcommand{\reqnomode}{\tagsleft@false\let\veqno\@@eqno}
\makeatother

\DeclareMathOperator{\Diff}{Diff}
\DeclareMathOperator{\Aut}{Aut}

\newtheorem{thm}{Theorem}

\newtheorem{prop}[thm]{Proposition}
\theoremstyle{definition}

\makeatother

\usepackage{float}

\begin{document}


\title{Reassessing the foundations of Metric-Affine Gravity}

\author{J. \textsc{François} $\,{}^{a,\,b,\,c,\,*}$ \and L. \textsc{Ravera} $\,{}^{d,\,e,\,f,\,\star}$ }

\date{}

\maketitle
\begin{center}
\vskip -0.3cm
\noindent
{\footnotesize{
${}^a$ Department of Mathematics \& Statistics, Masaryk University -- MUNI.\\
Kotlářská 267/2, Veveří, Brno, Czech Republic.\\[2mm]
 
${}^b$  Department of Philosophy -- University of Graz. \\
 Heinrichstraße 26/5, 8010 Graz, Austria.\\[2mm]
 
${}^c$ Department of Physics, Mons University -- UMONS.\\
20 Place du Parc, 7000 Mons, Belgium.
\\[2mm]

${}^d$ DISAT, Politecnico di Torino -- PoliTo. \\
Corso Duca degli Abruzzi 24, 10129 Torino, Italy. \\[2mm]

${}^e$ Istituto Nazionale di Fisica Nucleare, Section of Torino -- INFN. \\
Via P. Giuria 1, 10125 Torino, Italy. \\[2mm]

${}^f$ \emph{Grupo de Investigación en Física Teórica} -- GIFT. \\
Universidad Cat\'{o}lica De La Sant\'{i}sima Concepci\'{o}n, Concepción, Chile. \\[2mm]

\vspace{1mm}

${}^*$ {\small{jordan.francois@uni-graz.at}} \qquad \quad ${}^\star$ {\small{lucrezia.ravera@polito.it}}
}}
\end{center}



\vspace{0.5mm}

\begin{abstract}

We reassess foundational aspects of Metric-Affine Gravity (MAG) in light of the Dressing Field Method, a tool allowing to systematically build gauge-invariant field variables.
To get MAG started, one has to deal with the problem of ``gauge translations".
We first recall that Cartan geometry is the proper mathematical foundation for gauge theories of gravity, and that this problem never arises in that framework, which still allows to clarify the geometric status of gauge translations.
Then, we show how the MAG kinematics is obtained via dressing in a technically streamlined way, which highlights that it  reduces to a Cartan-geometric kinematics.


\end{abstract}

\vspace{0.2cm}


\tableofcontents

\bigskip

\clearpage

\section{Introduction}\label{Introduction}  
In the mid 70s was established \cite{Wu-Yang1975} the now well-known fact that the mathematical foundation of Gauge Field Theory (GFT), à la Yang-Mills (YM), is the differential geometry of Ehresmann (or principal) connections on fiber bundles \cite{Hamilton2018, Francois2021_II}. 
General Relativity (GR) already being inherently a geometric theory, in the late 70s and 80s, treatments stressing the geometric structure of all fundamental interactions appeared, e.g. \cite{Trautman, Eguchi-et-al980, GockSchuck}. 

The field of gauge theoretic formulations of gravity has its roots in Einstein's introduction, in 1925, of the local Lorentz group  $\SO(1,3)$ and vielbein ${e^a}_\mu$ \cite{Unzicker-Case2005, Sauer2006}, and in Weyl's introduction of the local spin group $\SL(2, \CC)$ (acting on spinors) in the 1929 paper \cite{Weyl1929} in which he also introduces for the first time the Gauge Principle for $U(1)$ \cite{ORaif1997}. 
But it started in earnest with Utiyama,  who, in 1955, introduced (independently from the simple $\SU(2)$ model published in 1954 by Yang and Mills \cite{Yang-Mills1954})  the general framework of GFT based on the general Gauge Principle for (i.e. of ``gauging" of) an arbitrary Lie group $G$, and thus applied it to the Lorentz group $S\!O(1,3)$ to recover the structure of GR \cite{Utiyama1956}. 
In the early 60s, Kibble \cite{Kibble1961} and Sciama \cite{Sciama1964} (re-)introduced the Einstein-Cartan formulation, which allows for spacetime torsion sourced by spinor fields.
Around that time, elucidation of the gauge structure of gravity was motivated, aside from deepening our understanding of it, by the expectations that it would facilitate its quantization -- as quantization of other GFTs has been successful, notably in the Standard Model (SM) -- and/or its unification with the other fundamental interactions. See e.g. \cite{Neeman:1998vkf}.
 The 70s thus saw a considerable activity in model building based on the gauging of various groups, and supersymmetrization thereof leading to supergravity models -- e.g. de Sitter groups by MacDowell and Mansouri \cite{McDowell-Mansouri1977}, or the conformal group \cite{Kaku_et_al1977}.
 
Gauge approaches to gravity have been dominated, understandably,  by the heuristic habits of Yang-Mills theory: 
Meaning that whenever one wanted to ``gauge" a Lie group $G$, one postulated a gauge potential $A$ with value in the Lie algebra $\LieG$ of $G$, defined by its gauge transformations under the gauge group, supposed to be $\G$; 
i.e. the group, under pointwise product, of maps $g : U \subset M \rarrow G$ with $M$ a (spacetime) manifold. 
Prominent examples of this are Poincaré gravity (PG), where $G=S\!O(1, 3)\ltimes \RR^4$ \cite{Hehl:2020hhp,Obukhov:2006gea,Mielke:2017nwt,Obukhov:2018bmf,Hehl:2023khc}, and its Metric-Affine gravity (MAG) generalization to $G=GL(n)\ltimes \RR^n$ with $n=$ dim$M$ \cite{Hehl-et-al1995,Hehl:1999sb,Vitagliano:2010sr,Percacci:2020bzf}. See    \cite{Blagojevi-et-al2013} for a bibliographic sample.
In mathematical, bundle geometric terms, it means that the gravitational gauge potential is seen as (the local representative of, see after) an Ehresmann connection on a $G$-principal fiber bundle $Q$ over  $M$, whose gauge group is $\G$. 
However, such approaches  feature systematically a group of ``internal gauge translations", conceptually redundant with the group of diffeomorphisms $\Diff(M)$, which gives rise to a set of issues that we shall review below. 
To even get started as theories of gravity, MAG and PG have to deal with these issues, which essentially implies to get rid of these gauge translations, or to
 try to identify them with $\Diff(M)$ (via the tetrad field/soldering form).

We have here two converging aims. 
First, we shall remind that a proper understanding of bundle geometry \mbox{forbids} such an identification (see our comment below eq. \eqref{SES-Q} giving the short exact sequence characterizing the bundle $Q$), but most importantly we will highlight that the proper mathematical foundation of gauge theories of gravity is \emph{Cartan geometry} \cite{Kobayashi1972, Sharpe, Cap-Slovak09, JTF-Ravera2024review}, 
i.e. the differential geometry of \emph{Cartan connection} \cite{Kobayashi1955, Marle} on fiber bundles, which captures the key Einsteinian insight about the nature of gravity and in which the issue of ``gauge translation" simply does not arise. 
Second, we shall expose and use the \emph{Dressing Field Method} (DFM) \cite{Francois2014,JTF-Ravera2024gRGFT,JTF-Ravera2024-SUSY,JTF-Ravera2025MISUDFM,JTF-Ravera2025offshellsusyDFM,JTF-Ravera2025bdyDFM}, a systematic approach to building gauge-invariants in GFT, to eliminate gauge translations, thereby streamlining the construction of the basic MAG and PG kinematics. 
Doing so will stress that the latter are actually just the kinematics one would obtain by starting from Cartan geometry in the first place.  

Considered together, these two goals result in our thesis that it is \emph{a priori} misguided to attempt to build a gauge theory of gravity by ``gauging" the  affine or Poincaré groups -- as this remains captive of the heuristic habits of YM theory, and overlooks the insight of Cartan geometry -- 
but also that even doing so, MAG and PG actually cannot be understood as genuine gauge theories stemming from applications of the Gauge Principle to these groups.
Indeed,~we shall stress that the tinkering done to make it work -- put on solid formal grounds via the DFM -- leads back to Cartan geometry, where one should have started from the beginning. 

\enlargethispage{.8\baselineskip}
The very same logic and conclusion apply e.g. to attempts to build theories of conformal gauge gravity by gauging the conformal group $S\!O(2,4)$, as well as for supersymmetrization of all the aforementioned theories.%
\footnote{
Relatedly, but distinct from supergravity model building,  conceptual confusion about which group to gauge and  misunderstanding of Cartan geometry are the root causes leading to obstacles in applying the supersymmetric  framework to obtain a general approach to a unified description of matter and interaction fields -- beyond the special $3D$ ``unconventional susy" (or AVZ) model, first proposing it \cite{Alvarez:2011gd, Alvarez:2021zhh, Alvarez:2013tga}. Attempts at building models of such a framework in $4D$ stumbled upon the issue of having to make sense of ``internal gauge translations" and to identify the translation potential with a separately postulated soldering form, leading to uncompelling efforts to justify a ``double metric structure".
These matters are addressed and solved via the DFM in \cite{JTF-Ravera2024ususyDFM, JTF-Ravera2025MISUDFM}.}

The  paper is thus structured as follows: In Section \ref{Mathematical conceptual commentary on gauge theories of gravity} we
 remind the bundle geometric structures underlying Yang-Mills type and gravitational gauge theories, emphasizing the distinct foundational role of Cartan geometry for the latter. 
Then, in Section \ref{The Dressing Field Method in a nutshell}, after briefly reviewing the gauge field-theoretic counterpart of the previously discussed global geometric structures, we present the key  technical and conceptual aspects of the DFM.  
All this lays the groundwork for Section \ref{Translations reduction in MAG via the DFM}, where MAG (and PG) kinematics is obtained via the DFM, thereby streamlining its technical and conceptual foundations: in particular, we automatically reproduce the so-called ``radius vector" usually introduced \emph{ad hoc} to handle the issue of gauge translations. 
Finally, we further discuss the implications of the DFM approach to MAG and PG in Section \ref{Discussion}, our main conclusion being that it only highlights that Cartan geometry is the sole sound foundation for gauge theories of gravity.

\section{Cartan geometric foundation of gauge theories of gravity}
\label{Mathematical conceptual commentary on gauge theories of gravity}

The underlying geometry of GFTs à la YM is that of a principal bundle $P$, a smooth manifold supporting the right action of a Lie group $H$ (its \emph{structure group}), $P\times H \rarrow P$, $(p, h) \mapsto ph\rdefeq R_h p$,  
whose orbits are the \emph{fibers}~of~$P$. 
The moduli space of fibers is itself a smooth manifold $M\defeq P/H$, so that there is a projection $\pi:P\rarrow M$, $p \mapsto \pi(p)=x$, s.t. $\pi(ph)=\pi(p)$. The bundle may be noted $P\rarrow M$. 
Its tangent bundle $TP$ contains thus the canonical vertical subbundle $VP\subset TP$, defined as the kernel of the tangent application $\pi_*: TP \rarrow TM$.
The linearization of the right action of $H$ defines the morphism of Lie algebras $ \LieH \rarrow VP$, ${\sf X} \mapsto {\sf X}^v$, with $\LieH$ the Lie algebra of $H$ and ${\sf X}^v$ a fundamental vertical vector induced by ${\sf X}\in \LieH$. 

The maximal group of automorphisms $\Aut(P)$ of $P$ is the subgroup of its  diffeomorphisms preserving its fiber structure -- $\Xi \in \Diff(P)$ s.t. $\Xi(ph)=\Xi(p)h$ -- i.e. mapping fibers to fibers.  
It thus induces by definition diffeomorphisms of $M$, so that there is a surjective group morphism $\Aut(P)\rarrow \Diff(M)$.
Its kernel is the normal subgroup $\Aut_v(P)$ of vertical automorphisms of $P$, i.e. those automorphisms acting only along fibers and inducing the identity transformation on $M$, i.e. $\id_M \in \Diff(M)$.
These are thus generated by maps $\upgamma: P \rarrow H$ with defining property $\upgamma(ph)=h\- \upgamma(p) h$,  forming,
under pointwise product, the \emph{gauge group} $\H(P)$ of $P$: 
There is then a isomorphism $\Aut_v(P)\simeq \H(P)$, given by $\Xi(p)=p\,\upgamma(p)$.\footnote{One may indeed check the defining automorphism property $\Xi(ph)=(ph)\, \upgamma(ph)=\ldots$, showing the defining property of $\upgamma$ to be essential. }
The geometry of $P$ is thus characterized by the short exact sequence (SES) of groups
\vspace{-3mm}
\begin{align}
\label{SES-P}
\id_P\rarrow \Aut_v(P) \simeq \H(P) \xrightarrow{\triangleleft} \Aut(P)  \xrightarrow{\t \pi}  \Diff(M) \rarrow  \id_M,
\end{align}
The local structure of $P$ is trivial, i.e. for $U\subset M$ it is the case that $P_{|U}\simeq U \times H$. 
Yet, in general 
$P\neq M\times H$.

As a manifold, $P$ has a de Rham complex of forms $\big(\Omega^\bullet(P), d \big)$, with $d$ the exterior derivative.
The equivariance of a form $\beta \in \Omega^\bullet(P)$  is defined by the pullback action of the structure group, $R^*_h \, \beta$. 
In particular,  given a representation $(\rho, V)$ of $H$, one defines \emph{equivariant} forms $\beta \in \Omega_\text{eq}^\bullet(P, V)$ whose equivariance is controlled by the representation: $R^*_h \, \beta=\rho(h)\-\beta$.
The gauge transformation of a form $\beta$ is defined by the pullback action by $\Aut_v(P)$, which, given the above isomorphism, is expressible in terms of the corresponding generating elements of $\H(P)$: 
$\beta^\upgamma \defeq \Xi^*\beta$. 
An~important case is that of  \emph{tensorial} forms $\alpha \in \Omega_\text{tens}^\bullet(P, V)$,  whose gauge transformations are homogeneous and controlled by the representation: 
$\alpha^\upgamma \defeq \Xi^*\alpha=\rho(\upgamma)\- \alpha$.
We remark that, in particular, gauge group elements are both equivariant 0-forms, $R^*_h\upgamma = h\- \upgamma h$, and tensorial 0-forms acting on each other by $\upeta^\upgamma \defeq \Xi^* \upeta = \upgamma\- \upeta \upgamma$. 
In~physics, tensorial 0-forms $\vphi \in \Omega_\text{tens}^0(P, V)$ represent matter fields.

Unfortunately, $d$ does not preserve tensorial forms, i.e. $d\alpha \notin \Omega_\text{tens}^\bullet(P, \rho)$.
To get a first order differential operator on $\Omega_\text{tens}^\bullet(P, \rho)$, one needs to introduce an \emph{Ehresmann connection} on $P$: that is 
$\omega \in \Omega^1_\text{eq}(P, \LieH)$, with defining properties
\begin{align}
 \label{def-prop-connection}
 \text{i) }\quad R^*_h \omega =\Ad(h\-)\, \omega =h\- \omega\, h, 
 \qquad \text{ and } \qquad 
\text{ii) }\quad \omega({\sf X}^v)={\sf X}.
\end{align}
\enlargethispage{1\baselineskip}
These properties implies that $\omega$  gauge transforms inhomogeneously: $\omega^\upgamma = \upgamma\- \omega \upgamma + \upgamma\- d\upgamma$. 
It also implies that it induces a covariant derivative on tensorial forms,  $D\_= d\_+ \rho_*(\omega)\_ : \Omega_\text{tens}^\bullet(P, V) \rarrow \Omega_\text{tens}^{\bullet+1}(P, V)$. 
This reflects the Gauge Principle, or gauge argument, of GFT. 
One shows that $D^2\alpha  = D\circ D\alpha=\rho_*(\Omega)\alpha$, where
$\Omega = d\omega+\tfrac{1}{2}[\omega, \omega]=d\omega+ \omega^2\, \in \Omega_\text{tens}^2(P, \LieH)$ is the \emph{curvature} of $\omega$.
It thus gauge transforms as $\Omega^\upgamma = \upgamma\- \Omega \upgamma$, and satisfies the Bianchi identity $D\Omega=d\Omega +\Ad_*(\omega)\Omega=d\Omega +[\omega, \Omega]=0$. 

The above, when restricted to $M$, or $U\subset M$, provides the complete kinematics of a YM gauge field theory, as we shall review briefly in next section, the only missing ingredient being a Lagrangian providing a specific dynamics.

\medskip

The mathematical underpinning of
 gauge theories of  gravity is  \emph{Cartan geometry}; see e.g. \cite{JTF-Ravera2024review} for a recent review, and \cite{Kobayashi1972, Sharpe, Cap-Slovak09} for in depth treatments, while 
 \cite{Kobayashi1957, Marle} are of historical interest.
 A Cartan geometry $(P, \b \omega)$ is  an $H$-principal bundle $P$ 
 endowed with a \emph{Cartan connection} 
 $\b \omega \in \Omega^1_\text{eq}(P, \LieG)$, with  $\LieG \supset \LieH$ s.t. $\LieG/\LieH\rdefeq \mathfrak p$ is a left $H$-module, satisfying the same  two defining properties \eqref{def-prop-connection} of an Ehresmann connection, but also a third distinctive one:
\begin{align}
\label{Key-prop-Cartan-geom}
    \text{iii) }\quad \b \omega : T P \rarrow \LieG\  \text{ is a linear isomorphism, i.e. } \ker(\b \omega)=\emptyset. 
\end{align}
From this single key property stems the specificities of Cartan geometry, which essentially boil down to the fact that  it ensures $P$ \emph{encodes the geometry of} $M$. 
Given the general-relativistic insight that gravity is the geometry of spacetime, Cartan geometry is thus perfectly adapted to describe the kinematics of gauge theories of gravitation, the Cartan connection $\b \omega$ representing a generalized gravitational gauge potential.

We emphasize that since a Cartan connection $\b\omega$ satisfies, like an Ehresmann connection $\omega$, the properties \eqref{def-prop-connection}, it transforms under the \emph{gauge group} $\H(P)$ as 
$\b\omega^\upgamma = \upgamma\- \b\omega \upgamma + \upgamma\- d\upgamma$: in other words, there is \emph{no gauge transformation ``associated to"} $\mathfrak{p}$. 
Since in many cases $\mathfrak{p}\simeq\RR^n$, there is  no ``internal gauge translations" in Cartan geometry. 

Technically, $\b\omega$  induces soldering on $M$,  i.e. the tangent bundle $TM$ is isomorphic to the associated vector bundle to $P$ with fiber $\mathfrak{p}$: we write $TM\simeq P\times_H  \mathfrak{p}$. In other words, vector fields on $M$ are  represented by $\Omega^0_\text{tens}(P, \mathfrak{p})$, and all tensors on $M$ are likewise represented by forms on $P$. 
Relatedly, given $\tau: \LieG \rarrow \mathfrak{p}$, a Cartan connection induces a soldering form $\theta \defeq \tau(\b\omega) \in \Omega^1_\text{tens}(P, \mathfrak{p})$, which thus $\H(P)$-transforms as $\theta^\upgamma =\upgamma\- \theta$. 
The curvature of $\b \omega$ is $\b{\Omega}\defeq d\b\omega +\tfrac{1}{2}[\b\omega, \b\omega]=d\b\omega+\b\omega^2 \in \Omega^2_\text{tens}(P, \LieG)$, and thus $\H(P)$-transforms as $\b{\Omega}^\upgamma = \upgamma\- \b\Omega\upgamma$.
It satisfies the Bianchi identity $\b D\b \Omega=d\b \omega + [\b\omega, \b\Omega]=0$. 
The \emph{torsion} of $\b\omega$ is $\Theta\defeq \tau(\b \Omega) \in \Omega^2_\text{tens}(P, \mathfrak{p})$, and $\H(P)$-transforms as $\Theta^\upgamma=\upgamma\-\Theta$. 
Manifestly, soldering and torsion are notions inexistent for Ehresmann connections, which on the upside accounts for the fact that Ehresmann geometry $(P, \omega)$ allows to describe an ``enriched structure" over $M$, unrelated to $M$'s intrinsic geometry, and is thus a perfect fit for YM type GFTs. See \cite{JTF-Ravera2024c} for a discussion of this point.

A Cartan connection is \emph{normal} if it is entirely expressed in term of its soldering part: $\b\omega_\text{\tiny N}=\b\omega_\text{\tiny N}(\theta)$. 
Typically this at least implies that $\Theta=0$. 
This generalizes the notion of Levi-Civita connection.
In reductive or parabolic Cartan geometries, one has a $H$-invariant decomposition $\LieG=\LieH \oplus \mathfrak{p}$, so $\b\omega = \omega+\theta$ where $\omega \in \Omega^1_\text{eq}(P, \LieH)$ is an Ehresman connection on $P$. 
Correspondingly, $\b\Omega =\Omega + \Theta$, but $\Omega\in \Omega^2_\text{tens}(P, \LieH)$ in general \emph{is not} $\omega$'s  curvature, yet contains~it. 

A \emph{flat} Cartan geometry $(P, \b \omega)$ is isomorphic to a Klein geometry $(G, \b\omega_\text{{\tiny MC}})$, where $G$ is a Lie group with Lie algebra $\LieG$ and closed subgroup $H$ so that it is an $H$-bundle over the homogeneous space ${\sf M}\defeq G/H$, i.e. \mbox{$G\rarrow {\sf M}$}.
The Maurer-Cartan form $\b\omega_\text{{\tiny MC}} \in \Omega^1_\text{eq}(G, \LieG)$ satisfies i)-iii) and
$d\b\omega_\text{{\tiny MC}}+ \tfrac{1}{2}[\b\omega_\text{{\tiny MC}}, \b\omega_\text{{\tiny MC}}]=0$; 
thus is a flat Cartan connection.
In~other words, Cartan flatness
implies that the manifold $M$ becomes (isomorphic to) a homogeneous space~${\sf M}$, 
 thus generalizing Riemann flatness which implies that $M$ becomes (isomorphic to) the  homogeneous space~${\sf M}=\RR^n$.

On $M$, or $U\subset M$, the local representative of $\b \omega$ and $\b \Omega$ are $\b A$ and $\b F$, which describe respectively the gravitational gauge potential and its field strength. 
The local representative of $\theta$ is $e\in \Omega^1(U, \mathfrak{p})$, which is none other than a ``vielbein".
Given~an $H$-invariant non-degenerate bilinear form $\eta:\mathfrak{p} \times \mathfrak{p}\rarrow \RR$, the Cartan connection thus induces a metric on $M$ by 
$g\circ e: \Gamma(TM)\times \Gamma(TM) \rarrow \RR$, $(\mathfrak{X}, \mathfrak{Y}) \mapsto g(\mathfrak{X}, \mathfrak{Y})\defeq \eta\big( e(\mathfrak{X}), e(\mathfrak{Y})\big)$.

\medskip
There is a relation between Cartan and Ehresmann geometry. 
Suppose  the $H$-bundle $P$ of the Cartan geometry $(P, \b\omega)$ can be embedded as a subbundle of a bundle $Q\rarrow M$ with structure group $G\supset H$, 
$\iota:P \hookrightarrow G$ ,  and with SES

\vspace{-3mm}
\begin{align}
\label{SES-Q}
\id_Q\rarrow \Aut_v(Q) \simeq \G(Q) \xrightarrow{\triangleleft} \Aut(Q)  \xrightarrow{\t \pi}  \Diff(M) \rarrow  \id_M,
\end{align}
where $\G(Q)$ is the gauge group of $Q$, whose elements are maps $\b\upgamma: Q \rarrow G$ with defining property $\b\upgamma(pg)=g\- \b\upgamma(p) g$. 
Observe that it contains gauge transformations ``associated to" $\mathfrak{p}$, that is ``gauge translations" in case $\mathfrak{p}=\RR^n$. 
It~should be clear from the SES \eqref{SES-Q} that these can in no way be identified with $\Diff(M)$, contrary to what is often stated in the literature,\footnote{For example, early in the well-known review \cite{Hehl-et-al1995} one reads ``\emph{On the face of it, local diffeomorphisms can be considered as locally gauged translations [...]}".
And in footnote 12, it is claimed that ``\emph{the group [of gauge translations] is locally isomorphic to the group of active diffeomorphisms}".}
as the whole gauge group $\G(Q)$ maps to $\id_M$.
\enlargethispage{1\baselineskip}

An Ehresmann connection on $Q$ is $\varpi \in \Omega^1_\text{eq}(Q, \LieG)$ satisfying, \emph{mutatis mutandis}, the two defining properties~\eqref{def-prop-connection}. 
It~therefore transforms under the gauge group $\G(Q)$ as 
$\varpi^{\b\upgamma}= \b\upgamma\- \varpi\, \b\upgamma + \b\upgamma\-d\b\upgamma$, thus in particular under gauge transformations corresponding to $\mathfrak{p}$ (``gauge translations").
An Ehresmann connection $\varpi$  on $Q$ induces by restriction a Cartan connection $\b\omega\defeq \iota^* \varpi$ on $P$,  
\emph{provided} the condition  $\ker(\varpi) \cap \iota_*(TP) = \emptyset$ is met -- so that iii) holds on~$P$. 
Reciprocally, a Cartan connection $\b\omega$ on $P$ induces an Ehresmann connection $\varpi$ on $Q$ 
satisfying this condition.
See \cite{Sharpe}, appendix A.3. 
The space of Ehresmann connections on $Q$ thus contains the space of Cartan connections on~$P$.

Locally, on $U\subset M$, the local representatives of both $\varpi$ and $\b\omega$ are forms $\b A \in \Omega^1(U, \LieG)$, and the only way to distinguish them is by how they transform: 
$\b A$ is the local representative of $\varpi$ on $Q$ if it transforms under the local version $\G$ of $\G(Q)$, i.e. with maps $\b \gamma : U \rarrow G$, 
and it is the local representative of $\b\omega$ on $P$ if it transforms under the local version $\H$ of $\H(P)$, i.e. with maps $\gamma : U \rarrow H$ (see eq. \eqref{Gauge-group} below for a more precise statement).

Gauge theories with gauge group $\G$, as MAG and PG purport to be, are actually concerned with the geometry of the bundle $(Q, \varpi)$ characterized by the SES \eqref{SES-Q}, which is not a Cartan geometry and therefore not the proper framework for a gauge theory of gravity. 
Yet, as we shall demonstrate, MAG and PG are actually no such \mbox{theories}, as their kinematics involves the elimination of gauge translations, which we shall perform systematically via the Dressing Field Method presented next, thereby  ending up with a kinematics with gauge group $\H$ stemming from the Cartan geometry $(P, \b\omega)$, as befitting gauge theories of gravity.

\section{The Dressing Field Method of gauge symmetry reduction}
\label{The Dressing Field Method in a nutshell}

The DFM \cite{Francois2014,JTF-Ravera2024gRGFT,JTF-Ravera2024-SUSY,JTF-Ravera2025MISUDFM,JTF-Ravera2025offshellsusyDFM,JTF-Ravera2025bdyDFM} is a systematic tool to produce gauge-invariant variables out of the field space $\Phi$ of a theory with gauge group $\H$ whose action on $\Phi$ defines gauge transformations. 
Let us briefly review its key aspects, to better appreciate the content of what comes next.
We start with the local, field theoretic, version of the geometric structures  discussed above. 

Consider a gauge theory over an $n$-dimensional manifold $M$, or region $U\subset M$, based on a (finite-dimensional) Lie group $H$, the \emph{structure group} of the underlying principal bundle $P\rarrow M$, with Lie algebra  $\LieH$.
Its elementary variables are: 
A YM gauge potential 1-form 
$A={A}_\mu \, dx^{\,\mu} \in \Omega^1(U, \LieH)$, 
with field strength 2-form 
$F=dA + \tfrac{1}{2}[A, A]=dA + A^2 \in \Omega^2(U, \LieH)$. 
They are respectively the local representatives of an Ehresmann connection $\omega$ and its curvature $\Omega$ on $P$.
Alternatively (or in addition) if one is considering a gauge theory of gravity, a gravitational gauge potential 1-form is 
$\b A={\b A}_\mu \, dx^{\,\mu} \in \Omega^1(U, \LieG)$, 
with curvature 2-form 
$\b F=d{\b A} + \tfrac{1}{2}[\b A, \b A]=d\b A + {\b A}^2 \in \Omega^2(U, \LieG)$. 
They are respectively the local representatives of a Cartan connection $\b \omega$ and its curvature $\b \Omega$ on $P$. 
Often the potential splits as $\b A = A + e$, where $e={e^a}_\mu dx^{\,\mu} \in \Omega^1(U, \mathfrak{p})$ is a soldering form.  
Matter fields are 
$\phi \in \Omega^0 (U,V)$, with $V$ a representation space for $H$, i.e. $\rho:H \rarrow GL(V)$, and $\rho_*:\LieH \rarrow \mathfrak{gl}(V)$. 
Their minimal coupling to gauge potentials is given by the covariant derivative, $D\phi:=d\phi+ \rho_*(A)\phi \in \Omega^1 (U,V)$. 
One shows that $D^2\phi= \rho_*(F)\phi$.

These fields are acted upon by the (infinite-dimensional) gauge group of the theory: the set of $H$-valued functions $\gamma: U \rightarrow H$, $x \mapsto \gamma(x)$, with
point-wise group multiplication $(\gamma \gamma')(x)=\gamma(x) \gamma'(x)$, defined by
\begin{align}
\label{Gauge-group}
\H := \left\{ \gamma, \eta :U \rightarrow H\ |\  \eta^\gamma\defeq \gamma^{-1} \eta\gamma\, \right\}.
\end{align}
This action defines the gauge transformations of the fields,
\begin{equation}
\begin{aligned}
\label{GT-bare-fields}
A^\gamma:=\gamma^{-1} A \gamma + \gamma^{-1} d \gamma, 
\qquad 
\alpha^\gamma:=\uprho(\gamma)\-\alpha,
\end{aligned} 
\end{equation}
where we designate collectively $\alpha=\{F, \b F, e, \phi, D\phi\}$, which are all ``gauge-tensorial" and $\H$-transform respectively via the adjoint (field strength), left (soldering), and $\rho$ (matter fields and their covariant derivatives) representations, that we denote collectively $\uprho=\{\Ad, \ell, \rho \}$.

The Lagrangian of a theory is a top form $L(A, e, \phi) \in \Omega^n(U, \RR)$ required to be quasi-invariant under the action of $\H$, i.e. $L(A^\gamma, e^\gamma, \phi^\gamma)= L(A, e, \phi) + db(\gamma; A, e, \phi)$, so that the field equations $E(A, e, \phi)=0$ are $\H$-covariant: $E(A^\gamma, e^\gamma, \phi^\gamma) = \uprho(\gamma)\- E(A, e , \phi)=0$. 
\medskip

Now, consider a subgroup $K$ of the structure group $H$, $K \subseteq H$, to which corresponds the gauge subgroup $\K \subseteq \H$. 
A $\K$-\emph{dressing field}  is a map $u: U\subset M \rarrow K$, 
i.e. a \emph{$K$-valued field}, defined by its $\K$-gauge transformation: 
\begin{align}
\label{GT-dressing}
u^\kappa:=\kappa\- u, \quad \text{ for } \kappa \in \K.
\end{align} 
Given the existence of a $\K$-dressing field, one defines the $\K$-invariant \emph{dressed fields} 
\begin{align}
\label{dressed-fields}
A^u\defeq  u\- A u + u\- du, 
\qquad
\alpha^u\defeq \uprho(u)\-\alpha.
\end{align}  
In particular, a dressed field strength is $F^u=u\- Fu = dA^u +\tfrac{1}{2}[A^u, A^u]$, i.e it is the field strength of the dressed potential.
A dressed matter field is $\phi^u=\rho(u)\- \phi$, and  $(D\phi)^u=\rho(u)\-D\phi= d\phi^u +\rho_*(A^u)\phi^u \rdefeq D^u\phi^u$, i.e. the dressed covariant derivative is the minimal coupling of the dressed matter field with the dressed gauge potential. 

Noticing the formal similarity with the gauge-transformations \eqref{GT-bare-fields}, one sees the simplest case
of the DFM ``\emph{rule of thumb}": To obtain the dressing of an object (field or functional thereof), first compute  its gauge transformation, then substitute in the resulting expression the gauge parameter $\gamma$ with the dressing field $u$. The dressed object is $\K$-invariant by construction.  

Note, however, that the dressing field \emph{is not} an element of the gauge group, $u \notin \mathcal{K}$, as it can be immediately seen by comparing \eqref{Gauge-group} and \eqref{GT-dressing}.
This is a crucial fact of the DFM: Despite the formal analogy with \eqref{GT-bare-fields}, the dressed fields \eqref{dressed-fields} are not gauge transformations. Hence, $\{A^u, \alpha^u\}$ \emph{must not} be confused with a gauge-fixing of the bare variables $\{A, \alpha\}$. 
Contrary to a gauge-fixing, the dressing operation is not a map from $\Phi$ to itself, but from $\Phi$ to the space of dressed fields $\Phi^u$, only isomorphic to a subspace (a subbundle) of $\Phi$ -- cf. \cite{JTF-Ravera2024gRGFT,Berghofer-Francois2024} for details. 
Clearly, when $u$ is an $\H$-dressing field, s.t. $u^\gamma =\gamma\- u$, the dressed fields are  $\H$-invariant.
\smallskip

A key aspect of the DFM is that a dressing field should be extracted/built from the (bare) field content $\upphi=\{A, \alpha\}$ of the theory, i.e. $u=u[\upphi]$, so that $u[\upphi]^\kappa := u[\upphi^\kappa] = \kappa\- u[\upphi]$.
In such a case the dressed field 
$\upphi^{u[\upphi]}$ have a natural interpretation as relational variables: they encodes the gauge-invariant relations among the physical degrees of freedom (d.o.f.) embedded redundantly in the bare fields (among pure gauge modes). 
The relational aspect of the theory is encoded in the gauge symmetry of the bare theory, which is then said \emph{substantive}, see \cite{Francois2018,JTF-Ravera2024gRGFT}.
If a dressing is introduced by \emph{fiat}, as additional d.o.f., one is dealing with a new, \emph{distinct} theory: The dressing field is \emph{ad hoc}, not built from the original d.o.f.; the dressed fields thus cannot be interpreted as representing the physical, relational content of the original theory. 
The gauge symmetry of the new theory is said to be \emph{artificial} \cite{Francois2018}.\footnote{
This case encompasses the so-called Stueckelberg trick, whereby one implements a gauge symmetry in a theory via the introduction of extra d.o.f., the Stueckelberg fields: 
It is clear that the latter are \emph{ad hoc} dressing fields, and what we described above is the inverse procedure of a Stueckelberg trick (which is thus an ``undressing" operation) \cite{JTF-Ravera2024gRGFT}. More broadly, the DFM underlies the so-called ``Massive Yang-Mills" models and, as a framework, it encompasses both the electroweak model and Stueckelberg-type models  \cite{Francois2018,JTF-Ravera2024gRGFT}.}

\paragraph{Residual transformations}

Being $\K$-invariant, the dressed fields \eqref{dressed-fields} are expected to display residual transformations under what remains of the gauge group. 
For these to be well-defined, it must be that $K$ is a normal subgroup of $H$, $K \triangleleft H$, so that $H/K\rdefeq J$ is again a Lie group. 
Correspondingly then, $\K \triangleleft \H$ and $\J =\H/\K$ is a gauge subgroup of $\H$.
In this case, the dressed fields may exhibit well-defined residual $\J$-gauge transformations, which are named \emph{residual transformations of the 1st kind}.
Now, since the $\J$-transformations of the bare fields are known, as a special case of their $\H$-transformations given by \eqref{GT-bare-fields}, to find the residual $\J$-transformations of the $\K$-invariant dressed fields \eqref{dressed-fields} one only needs to determine that of the dressing field $u$.
An interesting case is given by the following

\begin{prop}
\label{Resid-1st-kind}
If a $\K$-dressing field transforms as $u^\eta =\eta \- u \, \eta\, $ for $\,\eta \in \J$, then the dressed fields \eqref{dressed-fields} are standard $\J$-gauge fields with $\J$-gauge transformations
\begin{align}
  \label{J-trsf-dressed-fields}
  (A^u)^\eta = \eta \- A^u \eta + \eta\-d\eta, \quad  (\alpha^u)^\eta = \uprho(\eta \-)\, \alpha^u.
\end{align} 
In particular, the dressed curvature transforms as $(F^u)^\eta = \eta \- F^u \eta$, and a dressed matter field and its dressed covariant derivative as 
$(\phi^u)^\eta = \rho(\eta \-)\, \phi^u$ 
and 
$(D\phi^u)^\eta = \rho(\eta \-)\, D\phi^u$. 
\end{prop} 

Dressed variables may also be subject to residual transformations stemming from a possible ``ambiguity" in the choice  of the dressing field: 
Two dressing fields $u$, $u'$ are \emph{a priori}  related by $u'=u\zeta$, for $\zeta: U\rarrow K$ a map s.t. $\zeta^\kappa=\zeta$. 
Under pointwise product, these maps form a group we denote $\mathfrak G$ and call the group of
 \emph{residual transformations of the 2nd kind}. 
Its action on a dressing we may denote $u^\zeta\defeq u\zeta$, so that, while it does not act on bare fields $\upphi^\zeta:=\upphi$, it does act naturally on dressed ones as $(\upphi^u)^\zeta:= \upphi^{u\zeta}$. 
Explicitly, we have
\begin{prop}
\label{Resid-2nd-kind}
If the $\K$-dressing field transforms as $u^\zeta = u\zeta$ for $\zeta \in \mathfrak G$, then the dressed fields \eqref{dressed-fields} $\mathfrak G$-transform as
\begin{align}
  \label{G-trsf-dressed-fields}
  (A^u)^\zeta = \zeta \- A^u \zeta + \zeta\-d\zeta, \quad  (\alpha^u)^\zeta = \uprho(\zeta \-)\, \alpha^u.
\end{align} 
With in particular,  $(F^u)^\zeta = \zeta \- F^u \zeta$, while  
$(\phi^u)^\zeta = \rho(\zeta \-)\, \phi^u$ 
and 
$(D\phi^u)^\zeta = \rho(\zeta \-)\, D\phi^u$. 
\end{prop} 
However, when $u=u[\upphi]$ is built from the fields, the constructive procedure may typically reduce $\mathfrak G$  to a   \emph{discrete} group, reflecting the finite choices among the  d.o.f. available.
This happens e.g. in the context of classical and quantum mechanics, where  residual transformations of the 2nd kind  \eqref{G-trsf-dressed-fields} are shown to encode \emph{physical reference frame covariance} \cite{JTF-Ravera2024NRrelQM}. 
No such restriction naturally exists for \emph{ad hoc} dressing fields: in that  case, the action of $\mathfrak G$  on $\upphi^u$ essentially just replicates the action of  $\K$ on $\upphi$, the two situation are isomorphic.

\paragraph{Dressed dynamics}

As is clear from what precedes, a dressing is an operation performed at the kinematical level, turning the bare kinematics into a dressed one. 
Regarding dynamics, there are two possibilities to consider.

First, suppose that, following a Gauge Principle for the gauge group $\H$, one had built an $\H$-quasi-invariant Lagrangian for the bare variables $L(A, e,\phi)$, with bare field equations $E(A, e,\phi)=0$, defining an $\H$-gauge theory.
Then, given a dressing field $u$, one defines the \emph{dressed Lagrangian} by $L(A^u,e^u, \phi^u)\defeq L(A, e,\phi) + db(u;A, e,\phi)$ -- again seen to be a case of the DFM rule of thumb --
from which derives dressed field equations $E(A^u, e^u,\phi^u)=0$, i.e. equations for the dressed fields, which 
have just the same functional expression as the bare field equations.
In~case Proposition \ref{Resid-1st-kind} obtains, $L(A^u,e^u, \phi^u)$ is $\J$-quasi-invariant, and the dressed field equations $E(A^u, e^u,\phi^u)=0$ are $\J$-covariant. 
The $\H$-theory is thus seen to be reduced to  a $\J$-gauge theory.

Alternatively, one may start from the dressed kinematics, considered as the only physically relevant one, and take advantage of the greater freedom afforded by following a Gauge Principle for the residual gauge subgroup $\J$. 
That is, one may define a $\J$-gauge theory by building a $\J$-quasi-invariant Lagrangian $L'(A^u,e^u, \phi^u)$ from which derive $\J$-covariant field equations for the dressed fields $E'(A^u, e^u,\phi^u)=0$. 
Such a theory, in general not $\H$-quasi-invariant,  obviously cannot be considered to follow from a standard application of the Gauge Principle  to~$H$. 

As~it turns out, this second strategy is the one followed by model building in MAG and PG, as we stress below. So, contrary to heuristic claims often encountered in the literature, 
neither MAG nor PG follow straightforwardly from gauging of the  affine and Poincaré groups.

\section{Reduction of gauge translations in MAG via the DFM}\label{Translations reduction in MAG via the DFM}

We shall now apply the DFM to MAG. 
What follows essentially applies \emph{mutatis mutandis} to PG, i.e. by simply substituting the general linear group $GL(n)$ by the Lorentz group $S\!O(1, 3)$ -- or its spin cover $S\!pin(1,3)\simeq S\!L(2, \CC)$.
Let us start with reminding its \emph{a  priori} kinematical setup, 
introducing a compact matrix notation.

MAG starts with the gauging of the 
affine group  $GL(n)\ltimes T^n$, whose elements are pairs $(\sf{G},\mathsf{t})$ with composition law  $(\sf{G},\sf{t})\cdot (\sf{G}',\sf{t}')=(\sf{G}\sf{G}',\sf{t}+\sf{G}\sf{t}')$.
The neutral element is $(\1, 0)$, so the
inverses are $(\sf{G},\mathsf{t})\-=(\sf{G}\-, -{\sf G}\- t)$. 
By~definition of a semidirect product group, the subgroup $T^n$ is normal in $GL(n)\ltimes T^n$, i.e. $({\sf G}, t)\- \cdot (\1, t')\cdot ({\sf G}, t) \in T^n$.
 This group can be embedded in $GL(n+1)$: 
we have the injective group morphism   
\begin{equation}
\begin{aligned}
    GL(n)\ltimes T^n \quad &\hookrightarrow \quad GL(n+1),\\
    ({\sf G},{\sf t}) \quad&\mapsto\quad {\sf g} = \begin{pmatrix}
        {\sf G} & {\sf t} \\
        0 & 1
    \end{pmatrix}, \\
    ({\sf G},{\sf t})\- \quad&\mapsto\quad {\sf g}\- = \begin{pmatrix}
        \,{\sf G}\- & -{\sf G}\-{\sf t}\phantom{.}  \\
        \!\!0 & 1
    \end{pmatrix},
\end{aligned}
\end{equation}
such that indeed the semidirect group law of the affine group is reproduced by simple matrix multiplication,
\begin{align}
    {\sf g} {\sf g}' = \begin{pmatrix}
        {\sf G} & {\sf t} \\
        0 & 1
    \end{pmatrix} \cdot \begin{pmatrix}
        {\sf G}' & {\sf t}' \\
        0 & 1
    \end{pmatrix} = \begin{pmatrix}
        \,{\sf G} {\sf G}' & {\sf G} {\sf t}' + {\sf t}\phantom{.} \\
        0 & 1
    \end{pmatrix} .
\end{align}
The associated gauge group is $\mathcal{GL}(n) \ltimes \T^n \defeq \{ \gamma : U \rarrow GL(n) \ltimes T^n\, |\, \ldots\, \}$, similarly to the general case \eqref{Gauge-group}. 
Abusing notations slightly, we shall write gauge elements as $\gamma=\begin{pmatrix}
        \sf{G} & \sf{t} \\
        0 & 1
    \end{pmatrix}$. 
So, the gauge transformation of  gauge elements is 
\begin{align}
 \label{GT-gauge-element}
 \eta^\gamma= \gamma\- \eta \gamma
 =
 \begin{pmatrix}
        \sf{G} & \sf{t} \\
        0 & 1
    \end{pmatrix}\- \cdot 
    \begin{pmatrix}
        \sf{G}' & \sf{t}' \\
        0 & 1
    \end{pmatrix} \cdot
    \begin{pmatrix}
        \sf{G} & \sf{t} \\
        0 & 1
    \end{pmatrix}
    = 
    \begin{pmatrix}
        \sf{G}\-\sf{G}' \sf{G} & \sf{G}\- \sf{t}' \\
        0 & 1
    \end{pmatrix},
\end{align}
which indicates in particular that the gauge elements ${\sf t}$ are $\GL(n)$-tensorial and $\T^n$-\emph{invariant}, $T^n$-valued function. 
Correspondingly, an~element of the gauge subgroup $\mathcal{GL}(n)$ is
\begin{align}
\label{GLntrelemmatrix}
\mathbb{G} = \begin{pmatrix}
        \sf{G} & 0 \\
        0 & 1
    \end{pmatrix} ,
\quad \text{ with inverse }\quad 
\mathbb{G}\- = \begin{pmatrix}
       \, \sf{G}\- &\!\! 0 \\
        \!\! \!0 & \!\!1
    \end{pmatrix},
\end{align}
while an element of the additive Abelian normal  gauge subgroup $\mathcal{T}^n$ is given by the upper triangular matrix
\begin{align}
\label{matrixT}
    \mathbb{T} = \begin{pmatrix}
        \1 & \sf{t} \\
        0 & 1
    \end{pmatrix} ,
\quad \text{ with inverse }\quad 
    \mathbb{T}\- = \begin{pmatrix}
        \1 & - \sf{t}\phantom{.} \\
        0 & 1
    \end{pmatrix}.
\end{align}

Using this matrix embedding, 
the gauge potential of MAG and its field strength are
\begin{align}
\label{MAG-fields}
    \b A =
    \begin{pmatrix}
        A & V \\
        0 & 0 
    \end{pmatrix},
    \quad \text{and}\quad
    \b F 
    = d \b A + {\b A}^2 
    =
    \begin{pmatrix}
        F & T \\
        0 & 0
    \end{pmatrix} 
    =
    \begin{pmatrix}
        d A + A^2 & dV + A V \\
        0 & 0
    \end{pmatrix},
\end{align}
where $V$ is the gauge potential of translations, $A$ that of general linear rotations, and $T$ is the ``torsion" 2-form~of~$A$.
As the potential $\b A$ is being seen as the local, field-theoretical representatives of an \emph{Ehresmann} connection on the $G=\big(GL(n)\ltimes T^n\big)$-bundle $Q$ over $M$ -- rather than of a  \emph{Cartan} connection on the $H=GL(n)$-bundle $P$ over $M$ -- 
the fields \eqref{MAG-fields}  transform under the gauge group $\GL(n)\ltimes \T^n$ as
\begin{align}
\label{GT-A-F}
    {\b A}^\gamma 
    &= \gamma\- \b A \gamma + \gamma\- d \gamma 
    =
    \begin{pmatrix}
        \sf{G} & \sf{t} \\
        0 & 1
    \end{pmatrix}\-
    \begin{pmatrix}
        A & V \\
        0 & 0
    \end{pmatrix}
    \begin{pmatrix}
        \sf{G} & \sf{t} \\
        0 & 1
    \end{pmatrix}
    +
    \begin{pmatrix}
        \sf{G} & \sf{t} \\
        0 & 1
    \end{pmatrix}\- 
    \begin{pmatrix}
        d\sf{G} & d\sf{t} \\
        0 & 1
    \end{pmatrix}
    =
    \begin{pmatrix}
        {\sf G}\- A {\sf G} + {\sf G}\- d{\sf G} &\, {\sf G}\- \big( V + D \sf{t} \big)\phantom{.} \\
        0 & 0
    \end{pmatrix}, \notag 
    \\[1mm]
    {\b F}^\gamma & = \gamma\- \b F \gamma
    =
    \begin{pmatrix}
        \sf{G} & \sf{t} \\
        0 & 1
    \end{pmatrix}\-
    \begin{pmatrix}
        F & T \\
        0 & 0
    \end{pmatrix}
    \begin{pmatrix}
        \sf{G} & \sf{t} \\
        0 & 0
    \end{pmatrix}
    =\begin{pmatrix}
        {\sf G}\- F {\sf G}  &\, {\sf G}\- \big( T + F \sf{t} \big)\phantom{.} \\
        0 & 0
    \end{pmatrix},
\end{align}
where $D{\sf{t}} \defeq  d {\sf{t}} + A \sf{t}$ is the covariant derivative of the   $\GL(n)$-tensorial and $\T^n$-{invariant} 
 $T^n$-valued function $\sf{t}$. 
In~particular, this specializes to give the transformation of the MAG potential and field strength  under ``internal gauge translations",
\begin{align}
\label{GT-T-fields}
    {\b A}^\mathbb{T} = \mathbb{T}\- \b A \mathbb{T} + \mathbb{T}\- d \mathbb{T}
    =
      \begin{pmatrix}
         A  & V + D \sf{t} \phantom{.} \\
        0 & 0
    \end{pmatrix}
    \quad \text{and}\quad 
     F^\mathbb{T} = \mathbb{T}\- F \mathbb{T}
     =
     \begin{pmatrix}
     F  &\,  T + F \sf{t} \phantom{.} \\
        0 & 0
    \end{pmatrix}.
\end{align}

We may also observe that the fundamental representation of the affine group is $R^n\simeq T^n$, the corresponding right action of the gauge group on $X\in \Omega^0(U, \RR^n)$ is $X \mapsto ({\sf G}, {\sf{t}})\- X = {\sf G}\-(X - \sf{t})$, or using the matrix embedding, 
\begin{align}
\label{GT-X}
\b X \defeq \begin{pmatrix} X \\ 1
\end{pmatrix} \quad \mapsto \quad
{\b X}^\gamma = \gamma\- \b X 
=\begin{pmatrix} {\sf G}\-(X - {\sf{t}}) \\ 1
\end{pmatrix}.
\end{align}
The covariant derivative induced by $\b A$ is thus, \vspace{-1mm}
\begin{align}
  \b D \b X = d \b X + \b A \b X 
  = \begin{pmatrix} DX + V \\ 1
\end{pmatrix}, 
\quad \text{ and s.t. }
\quad 
(\b D \b X)^\gamma ={\b D}^\gamma {\b X}^\gamma = \gamma\- \b D \b X,
\end{align}
~with $DX \defeq dX + AX$. One furthermore shows that $\b D^2 \b X = \b F \b X= \begin{pmatrix} F X + T \\ 1
\end{pmatrix}$. 
The object $\b X \in \Omega^0(U, \RR^{n+1})$ is the local representative of a tensorial 0-form on the $(GL(n)\ltimes T^n)$-bundle $Q\rarrow M$, and can equally well be understood as a 
the section of an associated bundle $P\times_{GL(n)\ltimes T^n} \RR^{n+1}$. 
Typically in gauge theory, these represent matter fields, or ``Higgs" fields if the Lagrangian of the theory features a  potential term $V(\b X)$.
A priori, one may attempt to see~$\b X$, i.e. $X$, as the spacetime velocity of a  material point particle on $M$. However, there is obstruction to such an interpretation.
\medskip

The very existence of ``internal" gauge translations is a problem. 
First, and most notably,
they are redundant conceptually with diffeomorphisms $\Diff(M)$. 
And yet, contrary to what is often claimed in the MAG and Poincaré gravity literature,\footnote{See again footnote 3: one finds attempts to justify such claims by heuristic arguments, e.g. in \cite{Hehl-et-al1995} that
``\emph{this view gains some additional justification from the fact that
the gravitational field is coupled to the energy-momentum tensor density, i.e. to the translational current.}"} they can bear no relation to them because, 
as we stressed at the end of Section \ref{Mathematical conceptual commentary on gauge theories of gravity} and  as the SES \eqref{SES-Q} makes clear,
$\T^n$ as a gauge subgroup acts trivially on $M$, i.e. induces the identity of $\Diff(M)$.

Then, $\T^n$ makes impossible to identify  the most basic objects in the fundamental representation 
$X\in \Omega^0(U, \RR^n)$ with (components of) vector fields $\mathfrak{X} \in \Gamma(TM)$ of $M$; it is indeed clear by \eqref{GT-X} that while the former are $\GL(n)$-tensorial, as expected from vector field components, they are not $\T^n$-invariant. 
Consequently, the covariant derivative $DX$ in $\b D\b X$ cannot be understood as the covariant derivative of a vector field on $M$ -- and $\b D\b X=0$ is not a geodesic equation in $M$.\footnote{Had we dealt with PG, e.g. in $n=4$ dimension, instead of MAG, we would have found relatedly that the object $\psi \in \Omega^0(U, \CC^2)$ in the spin cover of the fundamental representation cannot be understood as a spinor field on $M$,  for it lacks invariance under gauge translations: so fermionic matter is not naturally represented. Nor is its minimal coupling to gravity, which is not $D\psi \in \Omega^1(U, \CC^2)$.}
Idem for objects in the dual of the fundamental representation $X^* \in \Omega^0(U, \RR^{n*})$, which cannot be identified with (components of) covectors, 1-forms, on $M$. So that, in general,  tensorial objects built from tensoring  $X$ and $X^*$ are not related to tensors of $M$. 

Finally, gauge translations 
make impossible the identification of the translation potential $V \in \Omega^1(U, T^n)$ with a soldering form inducing a metric on $M$ (and consequently make unclear the relation between $T$ and a true torsion tensor on $M$): 
Indeed, given a $GL(n)$-invariant non-degenerate bilinear form $\eta: T^n \times T^n \rarrow \RR$, if one tries to define a metric as $g\defeq \eta \circ V : \Gamma(TM) \times \Gamma(TM) \rarrow \RR$, $\mathfrak{X}, \mathfrak{Y} \mapsto g(\mathfrak{X}, \mathfrak{Y})\defeq \eta\big(V(\mathfrak{X}), V(\mathfrak{Y}) \big)$, then this metric is not $\T^n$-invariant. 

Thus, to even get started with MAG as a  gravity theory, one must somehow get rid of the gauge subgroup $\T^n$.\footnote{That much is clear from \cite{Hehl-et-al1995} 
where there is indeed no mention of a Lagrangian or field equations for MAG of PG \emph{before} the issue of gauge translations is dealt with.} 
The~DFM provides just the systematic framework that allows to do so naturally.

\subsection{Dressed connection and curvature}

\medskip
A dressing field for the gauge subgroup $\T^n$ of  ``internal translations"  is a map $u: U\subset M \rightarrow T^n$ \emph{defined} by $u^\mathbb{T}= \mathbb{T}\- u$, for $\mathbb{T} \in \mathcal{T}^n$.
Using again the matrix embedding, we write
\begin{align}
\label{dressing-for-T}
    u := \begin{pmatrix}
        \1 & \xi \\
        0 & 1 
    \end{pmatrix} , 
    \ \,  
     \text{ with  $\xi \in \Omega^0(U, T^n)$, \quad s.t. }
     \quad
    u^\mathbb{T} = \mathbb{T}\- u 
    =
    \begin{pmatrix}
        \1 & - \sf{t}\phantom{.} \\
        0 & 1
    \end{pmatrix} \begin{pmatrix}
        \1 & \xi \\
        0 & 1 
    \end{pmatrix} 
    = 
    \begin{pmatrix}
        \1 & \xi - \sf{t}\phantom{.} \\
        0 & 1 
    \end{pmatrix} .
\end{align}
We stress that, without the matrix embedding, one might have defined the dressing for $\T^n$ directly by $\xi: U\rarrow T^n$ s.t.   $ \xi^{\,\sf{t}} = \xi - \sf{t}$, as an additive Abelian version of the general definition of a dressing field. 

Given  a dressing field as above, applying \eqref{dressed-fields}, we easily build the $\T^n$-invariant dressed potential
\begin{align}
\label{dressed-connection}
     {\b A}^u := u\- \b A u + u\- d u 
    = \begin{pmatrix}
        \1 & - \xi \\
        0 & 1 
    \end{pmatrix} \begin{pmatrix}
        A & V \\
        0 & 0 
    \end{pmatrix} \begin{pmatrix}
        \1 & \xi \\
        0 & 1 
    \end{pmatrix} + \begin{pmatrix}
        \1 & - \xi \\
        0 & 1 
    \end{pmatrix} \begin{pmatrix}
        0 & d \xi \\
        0 & 0 
    \end{pmatrix} 
    =
    \begin{pmatrix}
        A & V + D \xi\phantom{.}\\
        0 & 0
    \end{pmatrix} =: \begin{pmatrix}
        A & e\phantom{.}\\
        0 & 0
    \end{pmatrix} ,
\end{align}
where $D \xi := d \xi + A \xi$. 
Correspondingly, the $\T^n$-invariant dressed field strength, i.e. the field strength of ${\b A}^u$, is
\begin{align}
\label{dressed-curv}
{\b F}^u \defeq u\- \b F u
    = \begin{pmatrix}
        \1 & - \xi \\
        0 & 1 
    \end{pmatrix} \begin{pmatrix}
        {F} & {T} \\
        0 & 0
    \end{pmatrix} \begin{pmatrix}
        \1 & \xi \\
        0 & 1 
    \end{pmatrix} 
    =
    \begin{pmatrix}
        {F} & {T} + {F} \xi \\
        0 & 0
    \end{pmatrix} 
    \rdefeq
    \begin{pmatrix}
        {F} & \Theta \\
        0 & 0
    \end{pmatrix}.
\end{align}
Comparison of \eqref{dressed-connection}-\eqref{dressed-curv} with \eqref{GT-T-fields} illustrates the DFM rule of thumb. 
We observe that the $\T^n$-invariant dressed field $e\defeq V+ D\xi \in \Omega^1(U, T^n)$ in \eqref{dressed-connection} is  called the ``key relation" of MAG in \cite{Hehl-et-al1995}.\footnote{Comparable results have been achieved in \cite{Tseytlin1982} via the notion of ``nonlinear realisations/representations" (NR), as defined e.g. in \cite{Coleman-et-al1969, Coleman-et-al1969bis, Cho1978}. 
One key difference between the DFM and NR is that, while in the former a dressing field is group-valued, the ``dressing factor" in NR is coset-valued. 
Furthermore, the DFM having a clear bundle geometric formulation, a dressing field (like all geometric objects) has a clean geometric equivariance under the gauge subgroup being eliminated (i.e. a ``linear", covariant, transformation), while the ``non-linear" transformation of the coset-valued field of NR seems non-geometric. 
The closest bundle geometric result that seems to connect to NR (that \cite{Tseytlin1982} indeed appears to be hinting at) is the \emph{Bundle Reduction Theorem} (BRT) -- see e.g. \cite{Trautman, Sternberg, Kob-NomI, Kob-NomII}. As detailed in \cite{Francois2014}, the DFM and the BRT can coincide when the structure group of the bundle under consideration is a (semi-)direct product of two subgroups. Which is precisely the case here.} 
Finally, one can build the dressed 0-form and its dressed covariant derivative
\begin{align}
\label{dressed-X-DX}
  {\b X}^u = u\- \b X 
  =
  \begin{pmatrix} X - \xi \\ 1
\end{pmatrix}
  \rdefeq 
  \begin{pmatrix} X^\xi  \\ 1
\end{pmatrix}
\quad \text{ and } \quad 
{\b D}^u {\b X}^u = d {\b X}^u + A^u {\b X}^u 
=\begin{pmatrix} D X^\xi + e \\ 1
\end{pmatrix}, 
\end{align}
with $D X^\xi = dX^\xi + A X^\xi$. 
Now,  $X^\xi = X - \xi \in \Omega^0(U, \RR^n)$, being $\T^n$-invariant, is potentially identifiable as a vector field on $M$ \emph{if} it retains the correct $\GL(n)$-tensorial transformation of its bare counterpart $X$.
Similarly, the  $\T^n$-invariant form $e:=V + D \xi \in \Omega^1(U, T^n)$ is a soldering form, and  $\Theta =De = de + A e$ is a true torsion 2-form on $M$, only if both are $\GL(n)$-tensorial.
To ascertain these questions, we must assess the residual transformations of the 1st kind of  the above dressed fields.

\subsection{Residual $\mathcal{GL}(n)$ transformations}
\label{Residual GL(n) transformations}

After reducing the normal gauge subgroup $\T^n$ via dressing, we expect  residual transformations of the 1st kind under $\mathcal{GL}(n)$.
As per the general explanations of Section \ref{The Dressing Field Method in a nutshell}, as we already know the $\GL(n)$-transformations of the bare variables by \eqref{GT-A-F} and \eqref{GT-X}, we need only find that of the dressing field $u$.

By assumption, the dressing field $\xi$ is in the fundamental representation of $\GL(n)$, so that
$\xi \mapsto \xi^{\sf G} \defeq {\sf G}\- \xi$. 
 Using the matrix embedding, and \eqref{GLntrelemmatrix}, this is
\begin{align}
\label{GL-trsf-dressing-fields}
    u^\mathbb{G} =  \mathbb{G}\- u \mathbb{G} = \begin{pmatrix}
        \sf{G}\- & 0 \\
        0 & 1
    \end{pmatrix} \begin{pmatrix}
        \1 & \xi \\
        0 & 1
    \end{pmatrix}  \begin{pmatrix}
        \sf{G} & 0 \\
        0 & 1
    \end{pmatrix} = \begin{pmatrix}
        \1 & \sf{G}\- \xi \\
        0 & 1
    \end{pmatrix} .
\end{align}

This is a special case of Proposition \ref{Resid-1st-kind}, 
which allows us to immediately conclude that the dressed fields are standard $\GL(n)$-gauge fields, so that by \eqref{J-trsf-dressed-fields} we have:
\begin{align}
    ({\b A}^u)^\mathbb{G} &= \mathbb{G}\- {\b A}^u\mathbb{G} + \mathbb{G}\- d \mathbb{G}
    =
    \begin{pmatrix}
        {\sf G}\-  A {\sf G} + {\sf G}\- d{\sf G} &\, {\sf G}\- e\phantom{.} \\
        0 & 0
    \end{pmatrix},
    \qquad  
    ({\b F}^u)^\mathbb{G} = \mathbb{G}\- {\b F}^u \mathbb{G}
    =
     \begin{pmatrix}
        {\sf G}\- F {\sf G}  &\, {\sf G}\- \Theta\phantom{.} \\
        0 & 0
    \end{pmatrix},  \notag\\[1.5mm]
    \label{GL-trsf-dressed-fields}
    ({\b X}^u)^\mathbb{G} &= \mathbb{G}\- \b X^u 
    =
    \begin{pmatrix}
        {\sf G}\- X^\xi   \\ 1
    \end{pmatrix}, 
    \quad \text{and} \quad 
    (D^u{\b X}^u)^\mathbb{G} = \mathbb{G}\- \b D^u\b X^u 
    =
    \begin{pmatrix}
        {\sf G}\- ( DX^\xi + e)   \\ 1
    \end{pmatrix}. 
\end{align}
The full group of local transformations of the dressed MAG kinematics is thus $\Diff(M)\ltimes \GL(n)$. 
From \eqref{GL-trsf-dressed-fields} it is now clear that $X^\xi$ is indeed identifiable with a vector field $\mathfrak X$ of $M$, and can now represent the spacetime velocity of a point particle on $M$.
So, $DX^\xi$ in ${\b D}^u {\b X}^u$ is the covariant derivative of vector fields, and $DX^\xi=0$  describes a geodesic~on~$M$. 
Tensors obtained from tensoring $X^\xi$ and $X^{\xi*}$ are true tensors of $M$.

Furthermore, it is clear that ${\b A}^u$ is but the (local representative of a) Cartan connection associated to a Cartan-affine geometry, with curvature ${\b F}^u$, inducing via  $e\in\Omega^1(U, \T^n)$, a true soldering form on $M$, a gauge-invariant  metric on $M$ by   $g\defeq \eta \circ e : \Gamma(TM) \times \Gamma(TM) \rarrow \RR$, $\mathfrak{X}, \mathfrak{Y} \mapsto g(\mathfrak{X}, \mathfrak{Y})\defeq \eta\big(e(\mathfrak{X}), e(\mathfrak{Y}) \big)$. 
We~have now a good kinematics for a gauge theory of gravity: 
But it is just the local version of the Cartan-affine geometry $(P, \b \omega)$, with $P \rarrow M$ a $H=GL(n)$-principal bundle (i.e. the frame bundle of $M$), we would have started with had we heeded the insight of Cartan geometry.

\section{Discussion}
\label{Discussion} 

Several observations and comments are in order.
First, we did not yet say how the $\T^n$-dressing field $u$  \eqref{dressing-for-T} is to be found: in the DFM, the physical picture changes significantly depending if the dressing is field-dependent or not.

If it is introduced as a separate object from the bare $\GL(n)\ltimes \T^n$ kinematics, i.e. as extra d.o.f., then it is what in Section \ref{The Dressing Field Method in a nutshell} we called an \emph{ad hoc} dressing field. 
In that form it reproduces what is known as the  ``radius vector", e.g. mentioned early in \cite{Hehl-et-al1995} (and attributed to Trautman \cite{Trautman1973}), an object indeed
introduced in MAG to get rid of gauge translations $\T^n$. 
But according to the DFM, this means that 
in MAG (and PG), which is thus the bare $\GL(n)\ltimes \T^n$ kinematics supplemented by an \emph{ad hoc} dressing field $u$/radius vector, $\T^n$ is an \emph{artificial} gauge symmetry -- also called ``fake" gauge symmetry by \cite{Jackiw-Pi2015} --
with no physical signature. 

Only the residual $\GL(n)$ (or $\SO(1, 3)$ in PG) gauge group has physical significance and  is  thus \emph{substantive}.
As~a matter of fact, and as observed already at the end of Section \ref{The Dressing Field Method in a nutshell} (and footnote 7), model building in MAG (and PG) starts only after eliminations of gauge translations $\T^n$ and the Lagrangians are only required to be 
 (quasi-) invariant under the residual $\GL(n)$-transformations.\footnote{That much is hinted at early in \cite{Hehl-et-al1995} where we read e.g. that ``\emph{Only after a certain reduction, the translational connection and curvature are converted into coframe and torsion, respectively"}[...]", in line with our comments after \eqref{dressed-X-DX} and \eqref{GL-trsf-dressed-fields} above. 
 }
 Usually, no invariance property under $\T^n$ is required. 
 It is thus incorrect to claim that MAG, or PG, are built from ``gauging" the affine or Poincaré groups à la Yang-Mills.\footnote{For example, \cite{Hehl-et-al1995} insists on considering ``\emph{[Poincaré] gravitational theories from the point of view of a Yang-Mills like gauging of the Poincaré group.}", while its section 2.7 is entitled ``\emph{Metric-affine gauge theories: gauging the [affine group] [...]}". 
The preface of \cite{Blagojevi-et-al2013} states that ``\emph{If one applies the gauge-theoretical ideas to [the Poincaré group], one arrives at the Poincaré gauge theory of gravity (PG)}".
In~the forewords of \cite{Blagojevi-et-al2013}, 
Kibble states ``\emph{applying the gauge principle to [the] Poincaré-group symmetries leads most directly not precisely to Einstein’s general relativity, but to a variant, originally proposed by Élie Cartan, which 
[...]
uses a spacetime with torsion.}"} 
 As we just showed above, at best MAG (and PG) kinematics 
 is just the kinematics of Cartan-affine geometry.

Looking at it the other way around, we see that one may have started with a Cartan-affine kinematics, i.e. $\b A'$ and $\b F'$ supporting $\GL(n)$-gauge transformations like \eqref{GL-trsf-dressed-fields}.
Then, we may   enforce  an artificial ``gauge translations" group $\T^n$, acting trivially on $\b A'$ and $\b F'$, by introducing a Stueckelberg field $u\-: U \rarrow T^n$ (i.e. $-\xi$) s.t. $(u\-)^{\mathbb T} = u\- \mathbb T$, and then defining the fields $\b A \defeq u \b A' u\- + udu\-$ and $\b F \defeq u \b F' u\- $, transforming under $\T^n$ as \eqref{GT-T-fields}: 
i.e. we end-up with a $\GL(n)\ltimes \T^n$ kinematics where $\T^n$ is ``fake". 
Clearly, $\b A'={\b A}^u$ and $\b F'={\b F}^u$;
as stressed earlier, the DFM encompasses  Stueckelberg tricks when dressing fields are \emph{ad hoc}.
\medskip

Furthermore, it could be argued that since in MAG/PG the dressing field is \emph{ad hoc}, according to Proposition~\ref{Resid-2nd-kind} and eq. \eqref{G-trsf-dressed-fields}, the dressed fields \eqref{dressed-connection} 
 -\eqref{dressed-X-DX}  may a priori support $\mathfrak G$-transformations of the 2nd kind, which all but reproduce the action \eqref{GT-T-fields}
 -\eqref{GT-X} of $\T^n$ on bare variables.

 A way out is to notice that in the field content $\upphi=\{\b A, \b F, \b X\}$ there is a natural candidate for a $\T^n$-dressing field: 
 we may indeed define 
 \begin{align}
 \label{X-dep-dressing}
     u=u[\b X]=\begin{pmatrix}
     \1 & X \\ 0 & 1
 \end{pmatrix}, \ \text{ which by \eqref{GT-X} is s.t. }\ 
 u[\b X]^{\mathbb T}\defeq& u[\b X^{\mathbb T}] ={\mathbb T}\- u[\b X], \\
 \text{ and }\
 u[\b X]^\mathbb{G}  \defeq& u[\b X^{\mathbb G}] = \mathbb{G}\- u[\b X] \mathbb{G}.
 \end{align}
Said otherwise, this is the field-dependent dressing field $\xi=\xi[\b X]=X$. 
It allows to define the $\T^n$-invariant  fields 
\begin{align}
\label{dressed-A-F-bis}
{\b A}^{u[\b X]} 
=
\begin{pmatrix}
        A & V + D X\phantom{.}\\
        0 & 0
\end{pmatrix} 
=:
\begin{pmatrix}
        A & e\phantom{.}\\
        0 & 0
\end{pmatrix} 
\quad \text{and} \quad
{\b F}^{u[\b X]} 
=
\begin{pmatrix}
        F & T + F X\phantom{.}\\
        0 & 0
\end{pmatrix} =: \begin{pmatrix}
        F & \Theta\phantom{.}\\
        0 & 0
\end{pmatrix} 
\end{align}
 by \eqref{dressed-connection}-\eqref{dressed-curv}, 
as well as 
\begin{align}
\label{dressed-X-DX-bis}
  {\b X}^{u[\b X]} = u[\b X]\- \b X 
  =
\begin{pmatrix} 0 \\ 1 \end{pmatrix}
\quad \text{ and } \quad 
({\b D} {\b X})^{u[\b X]} = d {\b X}^{u[\b X]} + A^{u[\b X]} {\b X}^{u[\b X]} 
=\begin{pmatrix}  e \\ 1
\end{pmatrix},
\end{align}
 similarly to \eqref{dressed-X-DX}.
Their residual $\GL(n)$-transformations are given by \eqref{GL-trsf-dressed-fields}.
The fields \eqref{dressed-A-F-bis}-\eqref{dressed-X-DX-bis} may be understood as relational variables encoding the $\T^n$-invariant relations among the ``internal translational" d.o.f. of $\b A$ and $\b X$. 
It~is indeed consistent with the a priori postulate of a $\GL(n)\ltimes \T^n$ kinematics, based on the bundled $Q\rarrow M$, that fields would have internal translational d.o.f. -- which  are still  entirely unrelated to $\Diff(M)$, as noted earlier. 
The object $\b X$ would then describe the ``generalised spacetime velocity" of a point particle with such  translational internal d.o.f. and  $\b X^{u[\b X]} =(0, 1)$ simply expresses that such a particle ``sees" itself at rest in its own reference~frame.\footnote{Something exactly analogous to \eqref{X-dep-dressing}-\eqref{dressed-X-DX-bis} can be done in conformal Cartan geometry, and conformal gravity, where a dressing field $u[\b Y]$ may be built from the dilaton  embedded in the \emph{tractor field} $\b Y \in \Omega^0(U, \RR^6)$ in the fundamental representation of $H \subset G= S\!O(2,4)$, and used to reduce Weyl gauge rescalings; this allows in particular to produce Dirac spinors from twistors \cite{Francois2019}.}

 Pushing the idea a step further, suppose we have a collection of $N$ such point particles, so 
$\upphi=\{\b A, \b F, \b X_1, \ldots \b X_N \}$.
Then, clearly, we have $N$ choices to define a dressing field $u_i=u[\b X_i]$,  $i \in \{1, \ldots N\}$,  giving rise to dressed fields 
$\upphi^{u_i}=\{{\b A}^{u_i}, {\b F}^{u_i}, {\b X_1}^{u_i} \ldots {\b X_i}^{u_i} \ldots {\b X_N}^{u_i}\}$, 
which are the relational variables describing the $\T^n$-invariant relations among  internal d.o.f. within $\upphi$ \emph{as seen from} the frame of $\b X_i$. 
The change of particle perspective  -- i.e.  of ``physical" reference frame -- is encoded by residual transformations of the 2nd kind, where $\mathfrak G$ is the discrete group of elements $\zeta_{ij}$ s.t. $u_j=u_i \,\zeta_{ij}$, so that $\upphi^{u_j} =\uprho(\zeta_{ij})\- \upphi^{u_i}$. 
This is analogous to the application of the DFM in non-relativistic classical and quantum mechanics \cite{JTF-Ravera2024NRrelQM}.

This view is not so bad, it could be interesting if one was in the business of writing theories for the affine gauge group $\GL(n)\ltimes \T^n$, in which case one might expect empirical consequences to the presence of the symmetry $\T^n$. 
That could also be the case e.g. if, instead of seeing $\b X$ as a type of matter field, one was trying to interpret it as a sort of ``gravitational Higgs field", by embedding it into an invariant potential $V(\b X)$, possibly leading to Higgs-type mechanism.\footnote{The review \cite{Hehl-et-al1995}, borrowing from Trautman \cite{Trautman1973, Trautman}, suggests to understand $\xi$ as a ``generalized Higgs field". As we noted above, this terminology would be apt only if a potential term appears in the Lagrangian. 
We remark that \cite{Francois2019} did implement such an idea in the context of conformal Cartan gravity mentioned in footnote 10, treating the tractor field as a Higgs field embedded in a potential implementing a Lorentz $\SO(1,3)$ symmetry breaking mechanism.}
The issue,  as stated earlier, is that this is not what MAG/PG model building is usually about, since it starts only \emph{after} kinematical elimination of $\T^n$ via dressing, and is constrained only by (quasi-)invariance under residual $\GL(n)$-transformations. 
These approaches have no observable consequences associated to gauge translations, so one is only really concerned by the physics underpinned by Cartan geometry.

\medskip
So, even if superficially MAG and PG approaches seemed to be 
taking a road to gauge gravity distinct from Cartan geometry, by gauging the affine or Poincaré groups à la Yang-Mills -- implying to consider $\b A$ as the local representative of an Ehresmann connection on the $G$-bundle $Q \rarrow M$ --
the actual practice to get them started,  involving the reduction of the gauge translation group $\T^n$ via the DFM, circles back to Cartan geometry and only highlights it as the sole sound foundation of classical gauge theories of gravity.

\section*{Acknowledgments}  

J.F. is supported by the Austrian Science Fund (FWF), grant \mbox{[P 36542]}, 
and by the Czech Science Foundation (GAČR), grant GA24-10887S.
L.R. is supported by the 
GrIFOS research project, funded by the Ministry of University and Research (MUR, Ministero dell'Università e della Ricerca, Italy), PNRR Young Researchers funding program, MSCA Seal of Excellence (SoE), 
CUP E13C24003600006, ID SOE2024$\_$0000103, of which this paper is part.

{
\normalsize 
 \bibliography{Biblio11}
}

\end{document}